\documentstyle[11pt]{article}
\input epsf

\begin{document}
\setcounter{page}{1}
\begin{titlepage}
\hfill Preprint YeRPHI-1507(7)-98
\vspace{2cm}
\begin{center}

{\bf
$R$-Parity Violation and Scalar Leptons Production  at Lepton-
Antilepton  Colliders}\\
\vspace{5mm}
{\large R.A. Alanakyan}\\
\vspace{5mm}
{\em Theoretical Physics Department,
Yerevan Physics Institute,
Alikhanian Brothers St.2,

 Yerevan 375036, Armenia}\\
 {E-mail: alanak@lx2.yerphi.am}\\
\end{center}

\vspace{5mm}
\centerline{{\bf{Abstract}}}
In the framework of  models with $R$-parity violation, scalar neutrinos
and scalar leptons production in the processes
 $l_i^+l_j^-\rightarrow\tilde {\nu}_{k,L} Z^0$ and
 $l^+_il^-_j\rightarrow \tilde{l}^{\mp}_{k,L}W^{\pm}$
is considered.We also consider within the Minimal
 Supersymmetric Standard Model
 Higgs bosons production in the processes
 $\mu^+\mu^-\rightarrow H^0_iZ^0$, $\mu^+\mu^-\rightarrow
 H^{\pm}W^{\mp}$ .
\vspace{5mm}

\vfill
\centerline{{\bf{Yerevan Physics Institute}}}
\centerline{{\bf{Yerevan 1998}}}

\end{titlepage}

                 {\bf 1.Introduction}

As known, in theories with $R$-parity violation \cite{F}-\cite{D} supersymmetric
particles may be produced singly as a result of their
superpartners collision.In particularly, in
\cite{BGH}-\cite{KRSZ} has been suggested
scalar neutrino production virtually and in resonance at
$e^+e^-$collisions, in \cite{HW}
its production virtually and in resonance at $\mu^+e^-$
collisions \footnote {for references on $\mu^+e^-$ and $\mu^+\mu^-$ -colliders see
 \cite{HW} and \cite{P} respectively.}.

However, the scalar neutrino mass  is not fixed in
 theory and, thus, we don't know  which energies are
necessary for scalar neutrinos production in resonance.

That is why it is necessary to study associated scalar neutrino
production with gauge bosons.

In \cite{RA} scalar neutrino production with photon in
lepton-antilepton colliders has been
considered:
\begin{equation}
\label{AA1}
 l_i^+ l^-_j\rightarrow \tilde{\nu}_k \gamma,
\end{equation}

Here we study scalar neutrinos and charged leptons production
with $W^{\pm}$-boson and $Z^0$-boson :

\begin{equation}
\label{AA2}
 l^+_i l^-_j\rightarrow \tilde{\nu}_{k,L} Z^0,
\end{equation}

\begin{equation}
\label{AA3}
 l^+_i l^-_j\rightarrow \tilde{l}^{\mp}_{k,L} W^{\pm}.
\end{equation}

It is interesting to notice that the result received below for
the processes (2),(3) are
applicable to the tree processes of
 charged and pseudoscalar Higgs bosons production in assotiation
 with gauge bosons
\footnote{The tree processes of Higgs bosons 
 production in association with photons in $\mu^+\mu^-$-collisions
has been considered in \cite{RA1},\cite{LT},\cite{ABDR},\cite{RA}.}:
\begin{equation}
\label{AA4}
 \mu^+ \mu^-\rightarrow H^0_3 Z^0,
\end{equation}
\begin{equation}
\label{AA5}
 \mu^+ \mu^-\rightarrow H^{\pm} W^{\mp},
\end{equation}
in the framework of the Minimal Supersymmetric Standard Model(MSSM),
see \cite{HK,GH} and references therein) at large $\tan\beta$
limit.

We do not consider here the processes
$ \mu^+ \mu^-\rightarrow H^0_{1,2} Z^0$ because the main
contribution to this processes comes from $s$-channel $Z^0$-bosons
 exchange diagram of
the Fig.2, however at large $\tan\beta$ the $s$-channel $Z^0$-boson
 exchange contribution
is supressed as $\frac{1}{\tan\beta}$, 
whereas other diagrams of the Fig.2 are
enhanced as $\tan\beta$.The $t,u$-channel contribution to the process
$ \mu^+ \mu^-\rightarrow H^0 Z^0$ within the Standard Model has
been considered in \cite{LT}, for main $s$-channel $Z^0$-boson exchange
contribution see e.g.\cite{ERZ} references therein.

{\bf 2.Results}

Using $R$-parity violating interaction with electrons and
neutrinos:
 \begin{equation}
\label{AA6}
{\cal L}=
h_{ijk}(\bar{l_i}P_Ll_j\tilde{\nu}_k+\bar{l_i}P_L\nu_j\tilde{l}_k)+H.c.
\end{equation}
we obtain the
following amplitudes
of processes (2),(3):
\begin{equation}
\label{AA7}
M=\frac{gh_{ijk}}{\cos \theta_W} \bar{u}(k_1)_i\left(a_L \frac{\hat{k}_4
\hat{Z}}{t}-a_R \frac{\hat{Z}\hat{k}_4}
{u}- \frac{(k_4Z)}{s-m_{\tilde{\nu}}^2} \right) P_{L}u(k_2)_j.
\end{equation}
where $a_{L}=-\frac{1}{2}+\sin^2\theta_W, a_{R}=\sin^2\theta_W $,
\begin{equation}
\label{AA8}
M=\frac{gh_{ijk}}{\sqrt{2}} \bar{u}(k_1)\left( \frac{\hat{k}_4
\hat{W}}{t}+2 \frac{(k_4W)}{s-m_{\tilde{\nu}}^2+im_{\tilde{\nu}} \Gamma_{\tilde{\nu}}}\right) P_{L}u(k_2).
\end{equation}   
Here we neglect the lepton masses and use the following notations:
 $Z_\mu$,$W_\mu$
is the polarization 4-vector of the $Z^0$- and $W^{\pm}_L$- bosons,
 $s=(k_1+k_2)^2$,
 $t=(k_1-k_4)^2$,
 $u=(k_2-k_4)^2$,
 $m_{\tilde{l}},m_{\tilde{\nu}}$
are the  masses of the scalar lepton and scalar neutrino respectively,
$\Gamma_{\tilde{\nu}}$ is the width of the scalar
neutrino,in case of the process (3) we denote $h_{ijk}=\sum_p h_{ipk}V_{pj}$ (we take
into account that typically in SUSY
models if we neglect leptons masses, the masses of scalar
leptons of the all flavours are equal to each other).

For the differential cross sections of the processes (2),(3) we obtain
the following results:

\begin{eqnarray}
\label{AA9}
&& \frac{d\sigma( l^+_i l^-_j\rightarrow \tilde{\nu}_{k,L} Z^0) }{dt}=
\frac{\alpha h_{ijk}^2}{8 \sin^2 \theta_W \cos^2\theta_W s^2}
((\frac{a^2_L}{t^2}+ \frac{a^2_R}{u^2})
(tu-m^2_{\tilde{\nu}}m^2_Z)+\nonumber\\&&
+\frac{2a_La_R(t-m^2_{\tilde{\nu}})(u-m^2_{\tilde{\nu}})}{tu}
+\frac{sm^2_{\tilde{\nu}}}{(s-m_{\tilde{\nu}}^2)}(\frac{a_L}{t}-\frac{a_R}{u})+\nonumber\\&&
+(\frac{1}{8}m_Z^2-\frac{1}{2}m_{\tilde{\nu}}^2) \frac{s}{(s-m_{\tilde{\nu}}^2)^2}),
\end{eqnarray}
\begin{equation}
\label{AA10}
\frac{d\sigma( l^+_i l^-_j \rightarrow \tilde{l}^{\mp}_{k,L}
W^{\pm}) }{dt}= \frac{\alpha h_{ijk}^2}{8 \sin^2 \theta_W s^2}
(a\frac{1}{t}-m_W^2m_{\nu}^2 \frac{1}{t^2}+b),
\end{equation}
\begin{equation}
\label{AA11}
a=m_W^2+m_{\tilde{l} }^2-s-\frac{2sm^2_{\tilde{\nu} } }
{(s-m_{\tilde{\nu} })^2+\Gamma_{\tilde{\nu}}^2m^2_{\tilde{\nu}}}
{(s-m_{\tilde{\nu}})^2},
\end{equation}
\begin{equation}
\label{AA12}
b=-1-\frac{2m^2_{\tilde{l}}s}{(s-m_{\tilde{\nu}}^2)^2+m^2_{\tilde{\nu}}\Gamma^2_{\tilde{\nu}}}
+
\frac{s}{2m_W^2}
\mid
-1+\frac{(s-m_{\tilde{l}}^2-m^2_W)}{s-m^2_{\tilde{\nu}}+im_{\tilde{\nu}} \Gamma_{\tilde{\nu}}} \mid^2,
\end{equation}
\begin{equation}
\label{AA13}
t_-<t<t_+,
\end{equation}
where
\begin{equation}
\label{AA14}
t_{\pm}=\frac{m_{{\tilde{\nu}},\tilde{l}}^2+m_{Z,W}^2-s\pm
\sqrt{(m_{{\tilde{\nu}},\tilde{l}}^2+m_{Z,W}^2-s)^2-4m_{{\tilde{\nu}},\tilde{l}}^2m_{Z,W}^2}}{2}
\end{equation}
for the processes (2),(3) respectively.

After performing integration within the limits (14),(15) 
we obtain for the total
cross sections the following result:
\begin{eqnarray}
\label{AA15}
&& \sigma( l^+_i l^-_j\rightarrow \tilde{\nu}_{k,L} Z^0)=
\frac{\alpha h_{ijk}^2}{8 \sin^2 \theta_W\cos^2\theta_W s^2}
(((a^2_L+a^2_R)
(m^2_Z+m^2_{\tilde{\nu}}-s)+\nonumber\\&&
+\frac{4a_La_Rm^2_{\tilde{\nu}}(s-m^2_Z)}{(m^2_Z+m^2_{\tilde{\nu}}-s)}-
-\frac{1}{2}\frac{sm^2_{\tilde{\nu}}}{(s-m_{\tilde{\nu}}^2)}) \log
(\frac{t_+}{t_-})+\nonumber\\&& +(t_+-t_-)(2a_La_R-2(a^2_L+a^2_R)+\frac{1}{8}\frac{s
(m^2_Z-4m^2_{\tilde{\nu}})}{(s-m_{\tilde{\nu}}^2)^2})),
\end{eqnarray}
\begin{equation}
\label{AA16}
\sigma( l^+_i l^-_j\rightarrow \tilde{l}^{\mp}_{k,L} W^{\pm}) =
\frac{\alpha h_{ijk}^2}{16 \sin^2\theta_Ws^2}(a\log (\frac{t_+}{t_-})+(b-1)(t_+-t_-)).
\end{equation}

At $\sqrt{s} \gg
m_{\tilde{\nu}},m_{\tilde{l}},m_W $ the previous
 formulas are reduced and we have:
 \begin{equation}
\label{AA17}
\sigma( l^+_i l^-_j\rightarrow \tilde{\nu}_{k,L} Z^0 )
=\frac{\alpha h_{ijk}^2}{4 \sin^2 \theta_W \cos^2 \theta_W s}
((a^2_L+a^2_R)
\log(\frac{s}{m_{\tilde{\nu}} m_Z})+a_La_R-(a^2_L+a^2_R)).
\end{equation}
\begin{equation}
\label{AA18}
\sigma( l^+_i l^-_j\rightarrow \tilde{l}^{\mp}_{k,L} W^{\pm})
=\frac{\alpha (h_{ijk})^2}{8 \sin^2 \theta_W s}
( \log(\frac{s}{m_{\tilde{l}} m_W})-1).
\end{equation}

On Fig. 3,4 we present the number of events
$\tilde{\nu}_{k,L}Z^0$ and $\tilde{l}^{-}_{k,L}W^- $
per year for the processes (2),(3) versus $m_{\tilde{\nu}}$,
$m_{\tilde{\nu}}$ at fixed
$\sqrt{s}$, at yearly luminosity
$L=1000 fb^{-1}$.In our numerical results we suppose that scalar
neutrino predominantly decay into $ W^{\pm}\tilde{l^{\mp}}$ pairs
 (i.e. $\Gamma_{\tilde{\nu}}=\Gamma(\tilde{\nu} \rightarrow W\tilde{l})$).

{\bf 3.Comparision with other mechanisms of scalar leptons production}

The processes\cite{FF1}-\cite{BKT}:
\begin{equation}
\label{AA19}
  e^+e^-\rightarrow \tilde{\nu} \tilde{ \nu^*},\tilde{l^+}\tilde{l^-}
\end{equation}
which proceed through virtual neutral gauge bosons (and their
superpartners exchanges)
have  a larger cross
 sections than the processes (2),(3) however, it becomes
 kinematically allowed at energies
  $\sqrt{s}>2m_{\tilde{\nu}},(2m_{\tilde{l}})$,
whereas process (3),(4) is kinematically allowed at lower
energies   $\sqrt{s}>m_{\tilde{\nu}}+m_Z,m_{\tilde{l}}+m_W$.

At LHC  scalar leptons  may be also produced in
pairs (see \cite{BRP} and references therein), however with increasing masses of the scalar
leptons the cross sections are decreases faster than in case of
the reactions (2),(3).

{\bf 4.Higgs bosons production with gauge bosons}

Using Higgs bosons interactions with leptons (A4)-(A6) we obtain
in the large $\tan\beta$ limit that
the amplitudes of the processes (4),(5)
are different only by coefficient from amplitudes
of scalar lepton and neutrino production
with $W^{\pm}$-boson and $Z^0$-boson considered in formulas
(7),(8) above :
\begin{equation}
\label{AA20}
M( \mu^+ \mu^-\rightarrow H^{\pm} W^{\mp})=\frac{g^2}{2}\frac{m}{m_W}\tan\beta \bar{u}(k_1)
(\frac{\hat{k}_4 \hat{W}}{t}+2 \frac{(k_4W)}{s-m_H^2})
P_{L}u(k_2),
\end{equation}
\begin{equation}
\label{AA21}
M(\mu^+ \mu^-\rightarrow H^0_3 Z^0)=M_L-M_R,
\end{equation}
\begin{equation}
\label{AA22}
M(\mu^+ \mu^-\rightarrow H^0_{1,2} Z^0)=M_L+M_R+M_0,
\end{equation}
where
\begin{equation}
\label{AA23}
M_{L,R}=\frac{g^2}{2\cos\theta_W}\frac{m}{m_W}\tan\beta
\bar{u}(k_1)\left(a_{L,R} \frac{\hat{k}_4
\hat{Z}}{t}-a_{R,L} \frac{\hat{Z}\hat{k}_4}
{u}\mp \frac{(k_4Z)}{s-m_3^2} \right) P_{L}u(k_2),
\end{equation}
where $M_0$ is the contribution from $s$-chanel $Z^0$-bosons exchange.

Amplitudes $M_{L,R,0}$ do not interfere with each other
and for differential cross section of the process (4) we obtain:
\begin{eqnarray}
\label{AA24}
&& \frac{d\sigma( \mu^+ \mu^-\rightarrow H^0_3 Z^0) }{dt}=
\frac{\pi\alpha^2}{8 \sin^4 \theta_W \cos^2\theta_W s^2}
\frac{m^2}{m^2_W}\tan^2\beta
((a^2_L+a^2_R)\nonumber\\&&(\frac{1}{u^2}+\frac{1}{t^2})(tu-m^2_3m^2_Z)
+\frac{4a_La_R(t-m^2_3)(u-m^2_3)}{tu}
+\nonumber\\&&+\frac{sm^2_3}{s-m_3^2}(a_L+a_R)(\frac{1}{t}-\frac{1}{u}))
\end{eqnarray}
Differential cross section of the process (5) and total cross section of
 processes (4),(5) may be obtained 
 using formulas (10)-(18) by replacements:

$h_{ijk}\rightarrow\frac{gm}{\sqrt{2}m_W}\tan\beta$
,$m_{\tilde{\nu}}\rightarrow m_3,
m_{\tilde{l}}\rightarrow m_4$ and $\Gamma_{\tilde{\nu}}=0$.

For instance, far from threshold we obtain:
 \begin{equation}
\label{AA25}
\sigma( \mu^+ \mu^-\rightarrow H^0_3 Z^0 )
=\frac{\pi\alpha^2}{2\sin^4 \theta_W \cos^2 \theta_W s}
\frac{m^2}{m_W^2}\tan^2\beta
((a^2_L+a^2_R)
(\log(\frac{s}{m_3 m_Z})-1)+a_La_R).
\end{equation}
\begin{equation}
\label{AA26}
\sigma( \mu^+ \mu^-\rightarrow H^{\mp} W^{\pm})
=\frac{\pi\alpha^2}{4\sin^4 \theta_W s}\frac{m^2}{m^2_W}\tan^2\beta
( \log(\frac{s}{m_4 m_W})-1).
\end{equation}
At $m_4>m_t$, $m_3 \approx m_4$ with high accuracy and consequently
 for
the processes (4),(5) we can use numerical results depicted on the 
Fig.3,4 for $h_{ijk}=10^{-2}$ which are the same as
for the processes (4),(5) at
$\tan\beta=17.5$ (in accordance with the
above-mentioned replacements).

Besides the contribution to the processes (2),(3) from the tree
diagrams of the  Fig.1
there is also a contribution from the loops  with virtual $W^{\pm}$-and
$t$-quarks and with other heavy particles in various extensions
of the Standard Model such as contributions from squarks, charged Higgs
bosons, chargino.Some of such contributions
(the $ZW^{\mp}H^{\pm}$ vertex) have been calculated
previously, see \cite{K} and references therein.
On the Fig.5 we depicted some of the box
diagrams which also describe the loop contribution to the
process (5).Naively the loop contribution is of order $\sigma
\sim \frac{\alpha^4}{s}$, 
it is possible that various amplitudes from various contributions
 are partly compensate each other at some parameters
 (a similar situation takes place
in the loop contribution \cite{WY} to the process 
$ \mu^+ \mu^-\rightarrow H^0_i\gamma$) and at these
parameters the
cross section may be essentially lower and the tree contribution will
dominate over the loop contribution especially at $\tan\beta \gg 1$.

It must be noted also, at $\sqrt{s} \gg M$(where M- is the mass of the heaviest
 particles in loop )that the loop contribution
also decreases  faster than the tree contribution, because loop
integrals contain the additional degree of $\frac{M^2}{s}$.

Using result of the \cite{RA} for the process 
$ \mu^+ \mu^-\rightarrow H^0_i\gamma$:
\begin{equation}
\label{AA27}
 \sigma(\mu^+ \mu^-\rightarrow H^0_3\gamma)=\frac{\pi \alpha^2}{2sin^2
 \theta_W} \frac{m^2}{m_W^2}\tan^2\beta \frac{1}{s-m_H^2}((1+
 \frac{m_H^4}{s^2}) \log( \frac{s}{m^2_{\mu}})-2
 \frac{m_H^2}{s}),
\end{equation}
we see that near threshold the cross section of the tree process 
$ \mu^+ \mu^-\rightarrow H^0_i\gamma$ exceed the cross section
 of process (5),far from threshold, as seen from comparision of
 the (25)
with (27), the cross section of the process
 of the process $ \mu^+ \mu^-\rightarrow H^0_i\gamma$ exceed the
 cross section of the process (5) in several times.

 \setcounter{equation}{0}
\appendix{{\bf Appendix A}}

\renewcommand{\theequation}{A.\arabic{equation}}
\indent

In the MSSM, the Higgs sector
contains two doublets of Higgs bosons with opposite hypercharge
($Y= \pm 1$ ).

After spontaneous symmetry breaking the
following physical states appear:charged Higgs bosons
 $H^\pm$, and three neutral ones,
 $H^0_1, H^0_2, H^0_3 $.

At tree level  the masses of charged Higgs bosons ($m_4$) and
scalars $H^0_{1,2}$
 and an angle $\alpha$ (which described the mixing of scalar
 states) are being expressed through the mass of
 pseudoscalar $H^0_3$ and $\ tan\beta=\frac{v_2}{v_1}$ where
 $v_2$,$v_1$  are both doublets vacuum expectations by  following
 relationships:
 \begin{equation}
\label{A13}
m^2_4=m^2_{3}+m^2_W
\end{equation}

\begin{equation}
\label{A14}
m^2_{1,2}=1/2\left[m^2_{3}+m^2_Z\pm((m^2_{3}+m^2_Z)^2
-4 m^2_Z m^2_{3}\cos^2 2\beta)^{1/2}\right]
\end{equation}

\begin{equation}
\label{A15}
\tan2\alpha=\frac{m^2_{3}+m^2_Z}{m^2_{3}-m^2_Z} \tan2\beta.
\end{equation}

It follows from (A1) that MSSM guarantes the existence of, at
least, one light Higgs boson with
$m_{2}<m_{Z}$.

Interactions of the Higgs bosons with muons are described by
lagragnian:
\begin{eqnarray}
\label{A16}
&&{\cal L}=
i\frac{gm}{2m_W}\frac{\cos\alpha}{\cos\beta}\bar{\mu}\mu H_1^0+
i\frac{gm}{2m_W}
\frac{\sin\alpha}{\cos\beta}
\bar{\mu}\mu H_2^0+\nonumber\\&&+
\frac{gm}{2m_W}\tan\beta\bar{\mu}\gamma_5\mu H_3^0
+(i\frac{gm}{\sqrt{2}m_W}\tan\beta\bar{\mu}P_L\nu H^++h.c.)
\end{eqnarray}
At $\tan\beta \gg 1$ the mass relation (A2),(A3) and formula
(A4) are strongly reduced:

\begin{equation}
\label{A17}
m_{2}=m_{3}, m_{1}=m_Z,
\frac{\sin\alpha}{\cos\beta}
 =\tan \beta \gg \frac{\cos\alpha}{\cos\beta}
at\,\, m_{3}<m_Z,
\end{equation}

\begin{equation}
\label{A18}
m_{2}=m_{H_Z}, m_{1}=m_{3},\frac{\cos\alpha}{\cos\beta}
 =\tan \beta \gg \frac{\sin\alpha}{\cos\beta}
at \,\,  m_{3}>m_Z .
\end{equation}
It must be noted, that radiative
corrections \cite{OYY}-\cite{BF} can strongly change
relations (A1),(A2) however
in the large $\tan\beta$ limit and at $m_{3}<m_Z$  or at
$m_{3} \gg m_Z$ formulas (A1),(A5),(A6) hold approximately true
even after taking into account the radiative corrections.

{\bf Figures Cuption}

Fig.1 Diagrams corresponding to the processes (2),(3).

Fig.2 Tree diagrams corresponding to the processes (4),(5).

Fig.3 Number of events $\tilde{\nu}_{k,L} Z^0 $ per year ($\sigma L$)
(at yearly luminosity  $L=1000 fb^{-1}$) produced in reaction (2)
as a function of $\sqrt{s}$ with $h_{\mu\mu k}=10^{-2}$.Curves 1,2,3
correspond to the $m_{\tilde{\nu}}= 0.1,0.5,0.7 TeV$ respectively.

Fig.4 Number of events
$\tilde {\l^{\mp}}_{k,L} W^{\pm} $ per year ($\sigma L$)
(at yearly luminosity  $L=1000 fb^{-1}$) produced in reaction (3)
as a function of $\sqrt{s}$ with $h_{\mu\mu k}=10^{-2}$.Curves 1-5
correspond to the $m_{\tilde{\nu}}=m_{\tilde{l}}=100
GeV$;$m_{\tilde{\nu}}=m_{\tilde{l}}=300
GeV$;$m_{\tilde{\nu}}=m_{\tilde{l}}=700
GeV$;$m_{\tilde{\nu}}=500,m_{\tilde{l}}=300
GeV$;$m_{\tilde{\nu}}=700,m_{\tilde{l}}=300 GeV$; respectively.

Fig.5 Loop diagrams corresponding to the processes (4),(5).Shaded ring
correspond to the diagrams with $Z^0W^{\mp}H^{\pm}$ -vertexes.

\setcounter{figure}{0}
\newpage
\begin{figure}
\begin{center}
\epsfxsize=10.cm
\leavevmode\epsfbox{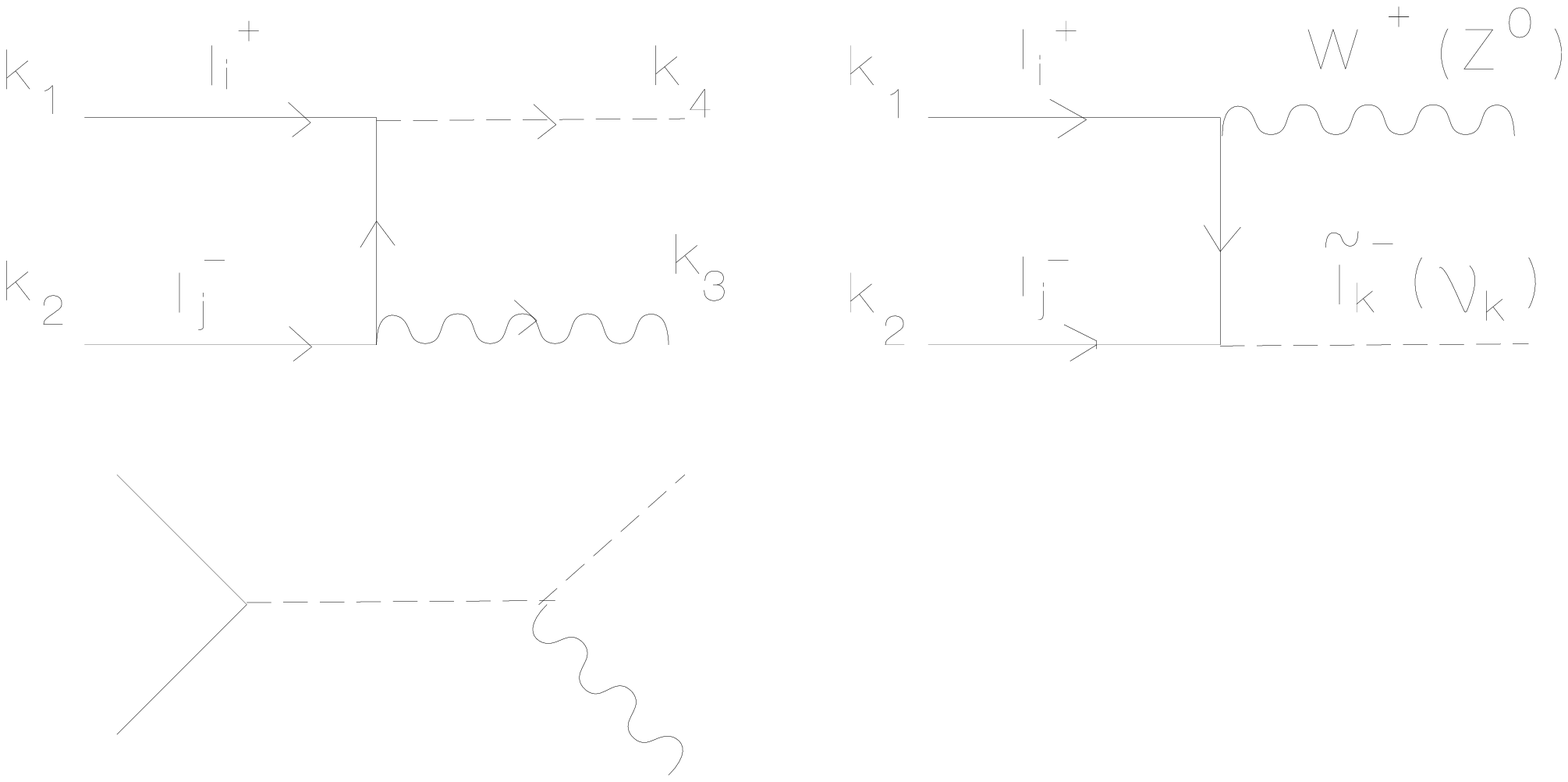}
\end{center}
\caption{}
\end{figure}

\begin{figure}
\begin{center}
\epsfxsize=10.cm
\leavevmode\epsfbox{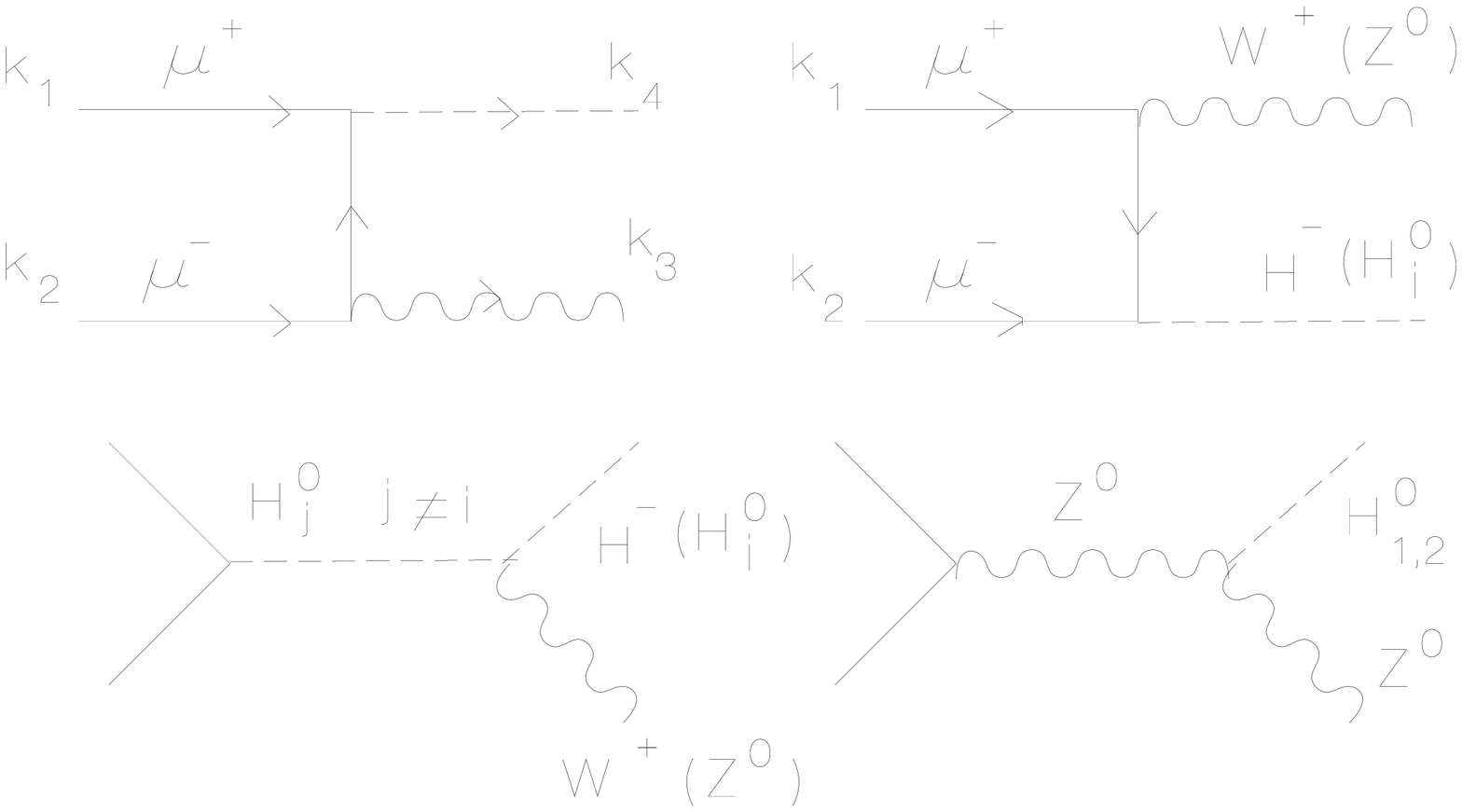}
\end{center}
\caption{}
\end{figure}

\newpage

\begin{figure}
\begin{center}
\epsfxsize=10.cm
\leavevmode\epsfbox{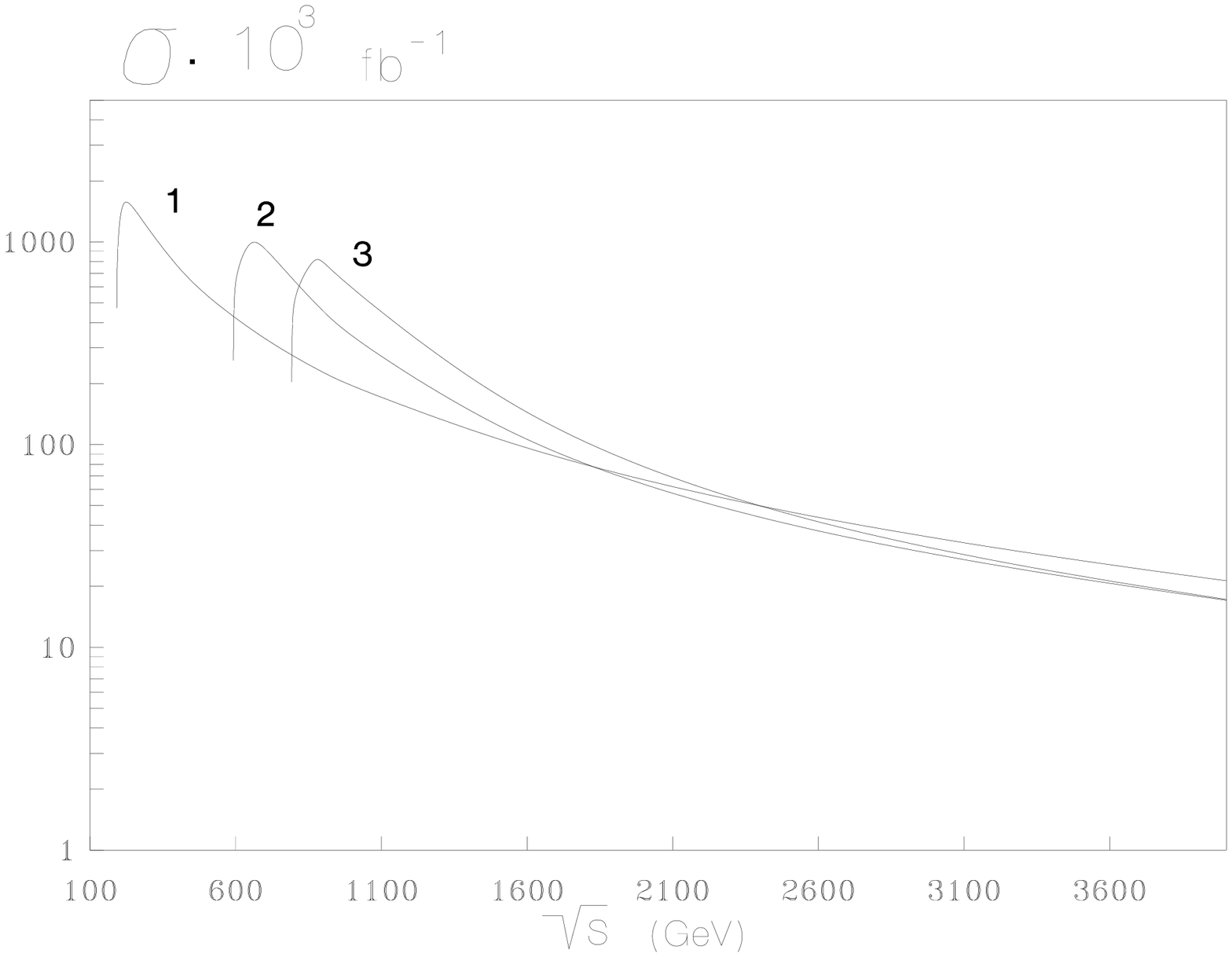}
\end{center}
\caption{}
\end{figure}
\newpage

\begin{figure}
\begin{center}
\epsfxsize=10.cm
\leavevmode\epsfbox{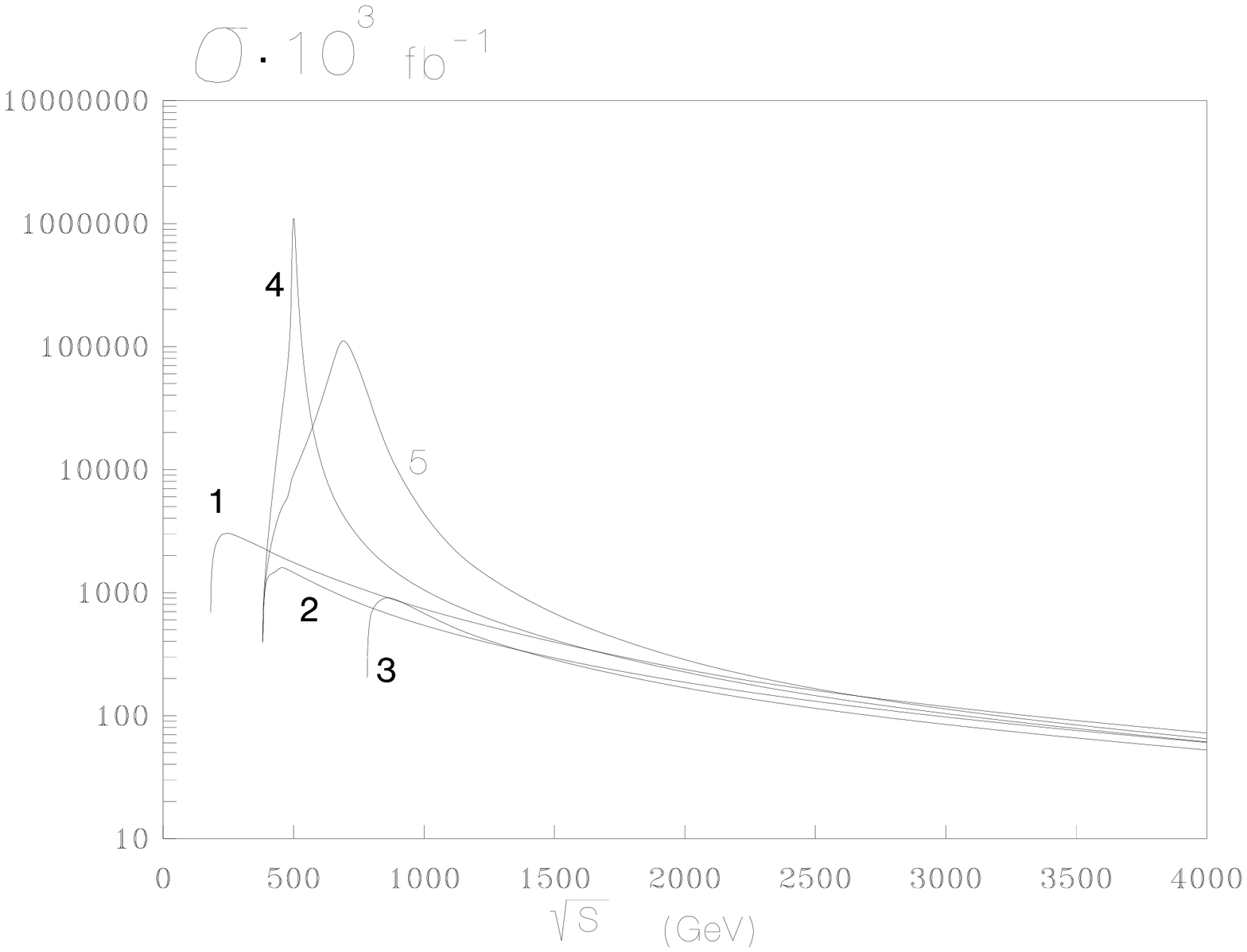}
\end{center}
\caption{}
\end{figure}
\newpage
\begin{figure}
\begin{center}
\epsfxsize=10.cm
\leavevmode\epsfbox{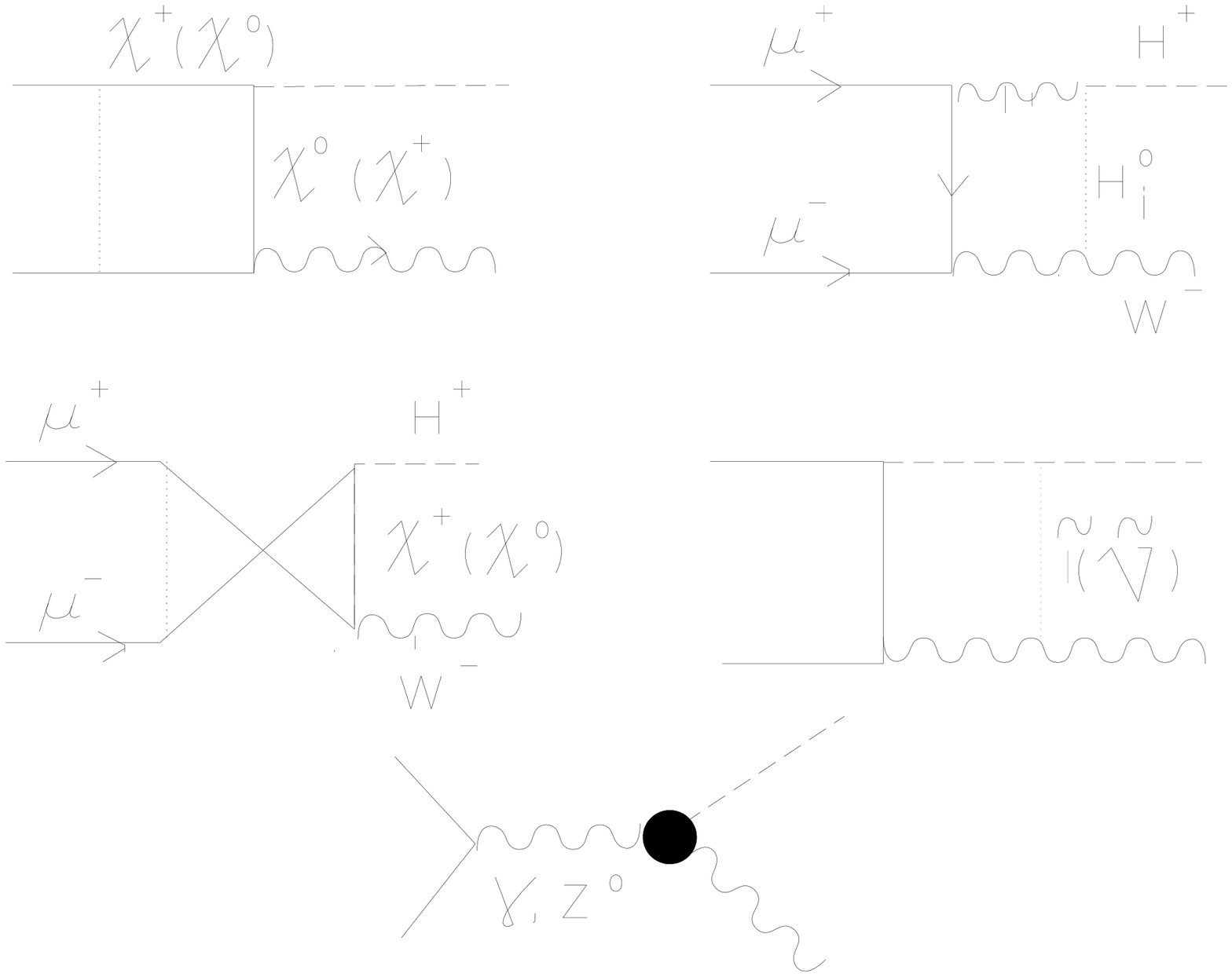}
\end{center}
\caption{}
\end{figure}

\end{document}